\documentclass[sigconf, nonacm]{stableranking-arxiv}

\usepackage{textcase}
\makeatletter
\expandafter\def\expandafter\@secfont\expandafter{\@secfont\MakeTextUppercase}
\makeatother
\setcopyright{none}
\date{}
\hypersetup{
  hidelinks,
  pdftitle={The General Stability of Ranking},
  pdfauthor={Houming Chen and H. V. Jagadish}
}

\usepackage{float}
\usepackage{comment}
\usepackage{needspace}

\usepackage{array}

\usepackage{graphicx}

\usepackage{tikz}
\usetikzlibrary{shapes.geometric,arrows.meta,positioning}

\usepackage{enumitem} 
\usepackage{multirow}
\usepackage{subfigure}

\usepackage[noend]{algorithm2e}
\RestyleAlgo{ruled}
\SetAlFnt{\small}
\SetKwComment{Comment}{/* }{ */}
\SetAlgoVlined
\LinesNumbered
\SetKwInOut{Hyperparameters}{Hyperparameters}
\SetKwInOut{Requirements}{Requirements}

\newtheorem{innercustomgeneric}{\customgenericname}
\providecommand{\customgenericname}{}
\newcommand{\newcustomtheorem}[2]{%
  \newenvironment{#1}[1]
  {%
   \renewcommand\customgenericname{#2}%
   \renewcommand\theinnercustomgeneric{##1}%
   \innercustomgeneric
  }
  {\endinnercustomgeneric}
}

\newcustomtheorem{customclaim}{Claim}
\newcustomtheorem{customthm}{Theorem}
\newcustomtheorem{customlemma}{Lemma}
\newcustomtheorem{customproposition}{Proposition}
\newcustomtheorem{customdefinition}{Definition}
\newcustomtheorem{customexample}{Example}
\newcustomtheorem{customproblem}{Problem}

\begin{document}
\raggedbottom
\title[The General Stability of Ranking]{The General Stability of Ranking}

\author{Houming Chen}
\affiliation{%
  \institution{University of Michigan}
  \city{Ann Arbor}
  \state{Michigan}
  \country{USA}
}
\email{houmingc@umich.edu}

\author{H. V. Jagadish}
\affiliation{%
  \institution{University of Michigan}
  \city{Ann Arbor}
  \state{Michigan}
  \country{USA}
}
\email{jag@umich.edu}

\begin{abstract}
Rankings derived from weighted scoring functions are widely used in university rankings and employment candidate evaluations. Since ranking
weights are often chosen by organizations or analysts, ranking stability asks
whether a reported ranking persists under reasonable weight changes. Prior work defines volume-based stability as the fraction of the weight space that induces the target ranking exactly.

This exact-match requirement can be too blunt: once a perturbed weight vector
produces a different ranking, exact stability gives it no credit, whether the
change replaces the top-ranked item or only swaps two nearly tied lower-ranked
items. We propose general stability, a distance-based generalization that
aggregates ranking regions according to a user-defined distance from the target
ranking. This lets users specify which ranking changes matter in the application,
while recovering exact stability as a special case.

  We extend existing methods and identify distance properties that permit faster  general-stability   computation with accuracy guarantees.
Our sampling algorithm, $GSC_{md}$, can take exponential time in the worst case, but a property we call \emph{uniform boundedness} gives a sample budget independent of database size and dimension at fixed accuracy and confidence.
We develop Prefix-GSC, a sampling optimization that reduces sorting for distances satisfying a further property, \emph{tail decay with bounded influence}.
We also develop  Conv-SC   to approximate stability through convex volumes for distances satisfying a property we term \emph{quasiconvexity}.

Experiments on seven real-data sources and generated inputs show how distance
affects robustness conclusions and when the algorithms save time.
On three 500-item university datasets, Prefix-GSC achieves
median query speedups of $1.22$--$1.98\times$ over $GSC_{md}$ at 2\% relative
error and 95\% confidence, including method setup.

\end{abstract}

\maketitle
\hypersetup{pdfauthor={Houming Chen and H. V. Jagadish}}

\thispagestyle{plain}
\pagestyle{plain}
\noindent{\small\textbf{Code and data:} \url{https://github.com/umich-dbgroup/General-Stability-of-Ranking}}\par\medskip

\section{Introduction} \label{sec:intro}

Rankings are widely used in settings such as university evaluation, job hiring, and public review websites. For instance, organizations such as U.S. News and Quacquarelli Symonds  (QS)  rank universities, and the resulting rankings can influence public perception and institutional decisions. Such rankings are commonly produced by scoring each item on several attributes, assigning weights to those attributes, and then sorting items by their weighted scores.

The choice of weights reflects a judgment about the relative importance of attributes, and is not driven by the data. Different reasonable choices of weights can lead to different rankings. This raises a robustness question: does the reported ranking reflect a stable pattern in the data, or does it depend on a narrowly limited choice of weights? If small changes in the weights lead to very different outcomes, the reported ranking provides weak support for a robust conclusion. It may also raise concerns that the reported weights might have been cherry-picked for a particular outcome.

Prior work formalizes this idea through volume-based stability~\cite{asudeh2018obtaining}, which measures the fractional volume of the weight space that induces a target ranking. We refer to this original notion as \textit{exact stability}. The following example illustrates the intuition: a ranking is robust when nearby weights support the same conclusion, and fragile when changes lead to different rankings.

\begin{figure}[tbp]
 \centering
  \includegraphics[height=0.36\linewidth,alt={Two-dimensional ranking regions for three candidates, showing a relatively large region where A is ranked above B above C.}]{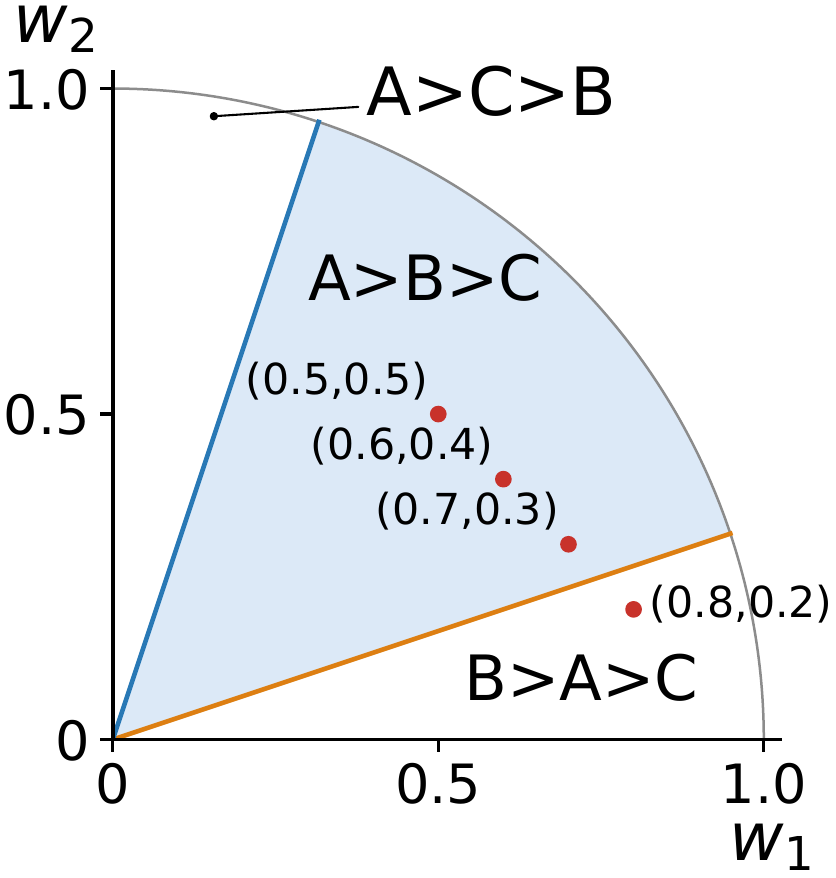}%
\hspace{0.004\linewidth}%
  \includegraphics[height=0.36\linewidth,alt={Two-dimensional ranking regions for six candidates, showing a thin exact region for A above B above C above D above E above F and nearby similar rankings.}]{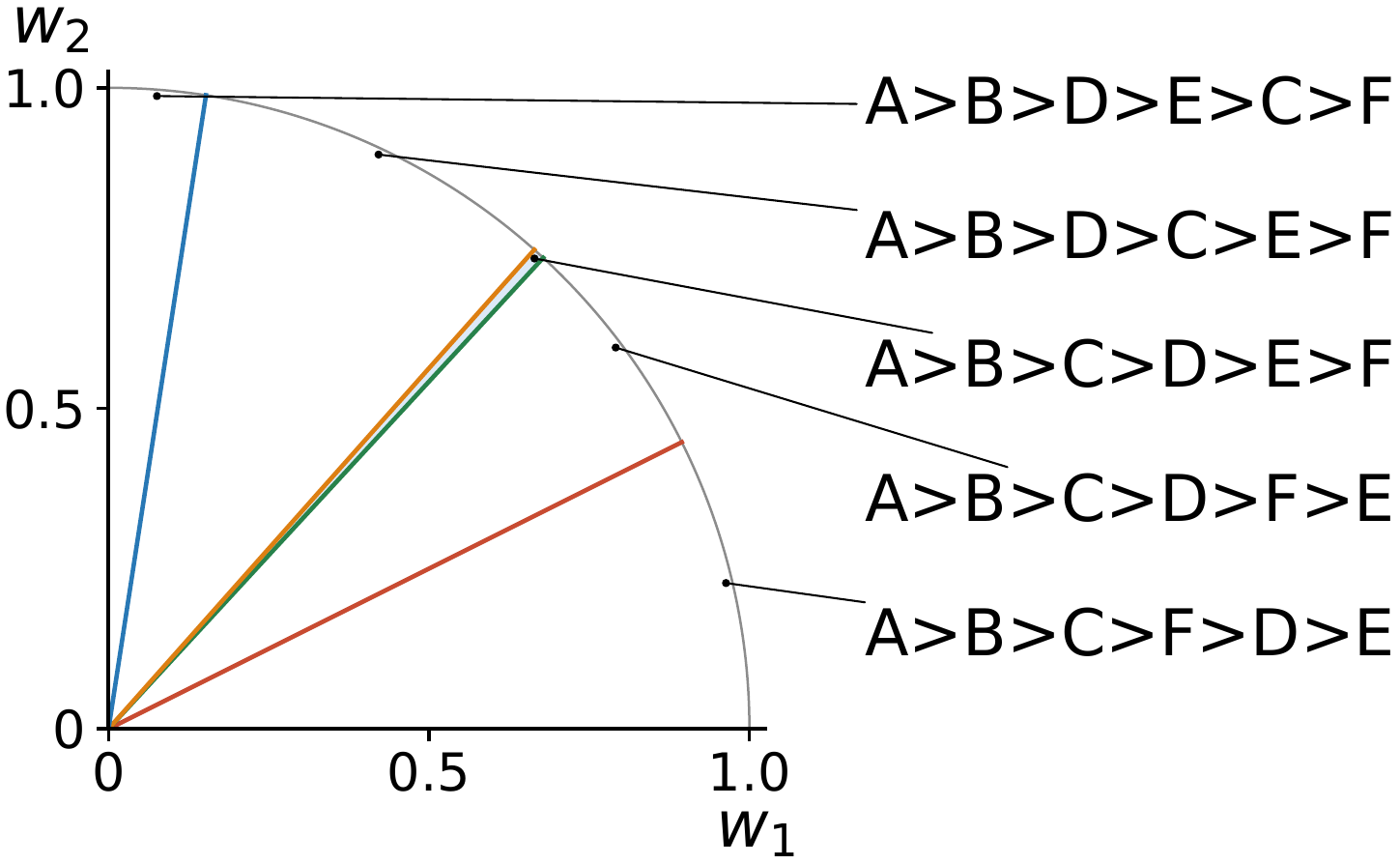}
\caption{Ranking regions for the motivating examples. Left: Example~\ref{Example:1}, where $A>B>C$ occupies a relatively large region. Right: Example~\ref{Example:2}, where $A>B>C>D>E>F$ has a small exact region but is adjacent to rankings that preserve the   winner and top-two order while changing only lower ranks.}
\Description{Two panels showing ranking regions in a two-dimensional weight space. The left panel shows a large region for A greater than B greater than C. The right panel shows a thin region for A greater than B greater than C greater than D greater than E greater than F next to similar ranking regions.}
\label{fig:intro_examples}
\end{figure}

\begin{customexample}{1} \label{Example:1}
Consider a hiring committee ranking candidates by a weighted sum of two criteria: an aptitude score $x_1$ and an experience score $x_2$. Suppose three candidates have scores shown in Table~\ref{tab:intro_example_1}.
\par\smallskip
\noindent\begin{minipage}{\linewidth}
\centering
\begin{tabular}{*7c} 
\toprule
 Cand. & $x_1$ & $x_2$ & \multicolumn{4}{c}{Weights and Corresponding Final Scores} \\ 
 \hline
  & &  & $(0.5,0.5)$ & $(0.6,0.4)$ & $(0.7,0.3)$ & $(0.8,0.2)$\\ 
 \midrule
 A & 97 & 95 & 96 & 96.2 & 96.4 & 96.6 \\ 
 B & 99 & 89 & 94 & 95 & 96 & 97 \\ 
 C & 84 & 94 & 89 & 88 & 87 & 86\\ 
 \bottomrule
\end{tabular}
\captionof{table}{The aptitude and experience scores of three candidates, and their final scores for different weight vectors.} \label{tab:intro_example_1}
\end{minipage}
\par\smallskip
A scoring function $f(x_1, x_2) = w_1x_1 + w_2x_2$ is used to compute the final scores. With weights $(w_1,w_2)=(0.6,0.4)$, the ranking is $A>B>C$. Nearby weights such as $(0.5,0.5)$ and $(0.7,0.3)$ induce the same ranking, while a larger change such as $(0.8,0.2)$ flips the top two candidates and produces $B>A>C$. In the left panel of Figure~\ref{fig:intro_examples}, the region inducing $A>B>C$ is relatively large; exact stability is precisely the volume fraction of this region in the weight space.
\end{customexample}

However, a small exact-stability region does not by itself tell us all we need to know. It may be that tiny changes in the weights lead to substantially different rankings, a serious robustness problem. Or, it may be that the nearby weight space is split among rankings that differ only by minor swaps. The distinction becomes clearer in the following example.
\begin{customexample}{2} \label{Example:2}
In the recruiting scenario described in Example~\ref{Example:1}, suppose we have six candidates with scores shown in Table~\ref{tab:intro_example_3}.

\par\smallskip
\noindent\begin{minipage}{\linewidth}
\centering
\begin{tabular}[t]{*3c}
\toprule
 Candidates & $x_1$ & $x_2$\\
 \midrule
 A & 100 & 98 \\
B & 90 & 97 \\
C & 90 & 86 \\
\bottomrule
\end{tabular} \quad
\begin{tabular}[t]{*3c}
\toprule
 Candidates & $x_1$ & $x_2$\\
\midrule
D & 81 & 94 \\
E & 77 & 88 \\
F & 90 & 76 \\
\bottomrule
\end{tabular}
\captionof{table}{The aptitude and experience scores of six candidates.} \label{tab:intro_example_3}
\end{minipage}
\par\smallskip
The ranking regions of these candidates are illustrated in the right panel of Figure~\ref{fig:intro_examples}. Consider the ranking $A>B>C>D>E>F$, which corresponds to the thin region in the middle of the figure. Under exact stability, this ranking is unstable because its own region has small volume. However, the neighboring regions do not drastically change the rankings. Moving the weights to one side gives $A>B>D>C>E>F$, which preserves the top two candidates and only swaps $C$ and $D$. Moving the weights to the other side gives $A>B>C>D>F>E$, which preserves the first four candidates and only swaps the last two. These perturbations are different from an unstable case where a small change replaces the leading candidate or substantially rearranges the top of the list. 
\end{customexample}

Large changes under small weight perturbations indicate fragility and may make the chosen weights appear cherry-picked. Small changes, however, can still support a reliable conclusion if much of the nearby weight space induces rankings that are not too dissimilar.

\textbf{From Exact Stability to General Stability.} The examples above motivate   extending  exact stability with a similarity-aware measure. We introduce \textit{general stability}, a distance-based generalization. Instead of measuring the volume of only the target region, general stability also incorporates the volumes of regions that induce rankings close to the target, with each region weighted according to a user-defined   ranking distance.

The   ranking distance  should be specified depending on which ranking changes matter in the application. For example, a user may care mostly about preserving the top-ranked items, or may want to distinguish a small local swap from a large rearrangement of the ranking. Different choices of   ranking distances  encode these judgments. If we assign distance $0$ to identical rankings and $\infty$ to all other rankings, general stability reduces to exact stability.

Prior measures of winner and top-$k$ stability~\cite{asudeh2018obtaining} are also special cases.
For winner stability, set the distance between two rankings to $0$ if they have the same top-ranked item, and $\infty$ otherwise.
For top-$k$-set stability, use distance $0$ if the rankings have the same top-$k$ items, and $\infty$ otherwise.
Ordered top-$k$ stability additionally requires these items to appear in the same order.

\textbf{Stability as a Robustness Measure.} It is important to clarify how stability should be interpreted. Stability is a robustness measure for a given ranking, not an objective that determines which ranking is best. A highly stable ranking is not necessarily the most appropriate ranking for an application; it   means that  reasonable changes in the weights  tend to produce similar rankings under the chosen distance. Therefore, the algorithmic task in this paper is to evaluate the robustness of a reported or user-specified ranking by computing its stability   value, rather than optimizing over rankings to maximize stability.

\textbf{Computing Stability.}   Exact computation is difficult for
exact stability. We extend the sampling method of
\citet{asudeh2018obtaining}  to general stability.
 The resulting algorithm, $GSC_{md}$, can still take exponential time in the worst case.
This raises a practical question: when can users obtain an accurate
stability estimate efficiently?

  We show that $GSC_{md}$ is efficient when the distance has a property we call \emph{uniform boundedness}.
We introduce Prefix-GSC, a sampling optimization that reduces sorting per sample for distances satisfying another property, \emph{tail decay with bounded influence}.
We also propose  Conv-SC, an algorithm that uses convex-volume approximation for distances  satisfying a property we term   \emph{quasiconvexity}.
These results help users choose a computation method for their distance.

 \subsection*{Summary of Our Contributions}

Our main contributions are as follows:

\begin{enumerate}
    \item We introduce and formalize   \textit{general stability}, which
    incorporates user-defined distance functions into stability analysis   and
    recovers exact stability as a special case.

    \item We study practical distance functions that distinguish small ranking
    changes from large ones and allow users to specify which parts of the
    ranking matter most.

    \item We extend prior exact-stability methods~\cite{asudeh2018obtaining}
    to compute general stability.

    \item We identify three distance properties that allow faster computation.
    Uniform boundedness makes $GSC_{md}$ efficient, and we develop Prefix-GSC
    and Conv-SC to exploit tail decay with bounded influence and quasiconvexity,
    respectively. Both methods have explicit error guarantees.

    \item   Our case studies show how the choice of distance changes conclusions
    about a ranking's robustness. Experiments on seven
    real-data sources and generated inputs demonstrate runtime gains at
    comparable accuracy and identify when the proposed methods are worthwhile.
 \end{enumerate}

\section{Problem Setup} \label{sec:2}

\subsection{Preliminaries} 

We consider a database \(\mathcal{D} = \{t_1, t_2, \ldots, t_n\}\) of \( n \) items. Each item \( t_i \in \mathcal{D} \) is described by a vector of \( d \) scoring attributes:
\[
t_i = \langle t_i[1],\ t_i[2],\ \ldots,\ t_i[d] \rangle .
\]
  To rank the items, we assign weights to the attributes and sort by their
weighted scores. Each weight expresses the importance of one attribute,
such as aptitude or experience in Example~\ref{Example:1}.

A ranking \(\mathfrak{r}\) is a bijective mapping from \(\mathcal{D}\) to \(\{1, 2, \ldots, n\}\). \(\mathfrak{r}(t_i)\) indicates the rank of \(t_i\), and \(\mathfrak{r}^{-1}(k)\) denotes the item in the \(k\)-th position. We focus on rankings induced by linear scoring functions.

\begin{customdefinition}{1}[Ranking Induced by a Linear Scoring Function]
A linear scoring function is a mapping \( f_{\mathbf{w}} : \mathbb{R}^d \to \mathbb{R} \) defined by
  $f_{\mathbf{w}}(t) \;=\; \sum_{i=1}^d w_i\, t[i],$
 where \(\mathbf{w} = \langle w_1, w_2, \ldots, w_d\rangle\) has \( w_i \geq 0 \) for all \( i \) and \(\|\mathbf{w}\|_2\le 1\). Equivalently, \(\mathbf{w}\) lies in the positive orthant of the closed unit ball in \(\mathbb{R}^d\), following the weight-space convention in prior stability work~\cite{asudeh2018obtaining}.  The ranking \(\nabla_{f_{\mathbf{w}}}(\mathcal{D})\) is obtained by ordering the items in \(\mathcal{D}\) in descending order of their scores \( f_{\mathbf{w}}(t)\).
 Throughout the paper, fixed item IDs break score ties.
 \end{customdefinition}

Let \(\mathcal{W}\) be the set of all feasible weight vectors.
  Positive rescaling of \(\mathbf w\) does not change the
ranking, so the ball represents the possible weight directions.
In some situations, the user might want to set up an acceptable region
$\mathcal{W}^* \subseteq \mathcal{W}$, which represents the set  of weight
vectors   considered acceptable for the application. We require this
region to be measurable and have positive volume. When no restriction is
stated, $\mathcal W^*=\mathcal W$.

\subsection{Ranking Regions} \label{sec:regions}

  Weight vectors producing  the same ranking  form a \emph{ranking region}.
  For a ranking \(\mathfrak{r}\) of \(\mathcal{D}\), this region is
$\mathcal{R}_{\mathcal{D}}(\mathfrak{r}) \;=\; \bigl\{\, \mathbf{w} \in \mathcal{W} \,\big|\; \nabla_{f_{\mathbf{w}}}(\mathcal{D}) = \mathfrak{r} \bigr\}.$
Within an acceptable region, write
$\mathcal R^*_{\mathcal D}(\mathfrak r)=
\mathcal R_{\mathcal D}(\mathfrak r)\cap\mathcal W^*$.

We now examine the geometry of these ranking regions. First, consider the 2D example from Example~\ref{Example:1}. There, \(\mathcal{W}\) (the unit quarter disk) is divided into sectors by rays through the origin, each sector corresponding to a unique ranking. Our goal is to generalize this observation to higher dimensions.

  Except for score ties, item \(t_i\) ranks before \(t_j\)  if and only if
  \(f_{\mathbf w}(t_i)>f_{\mathbf w}(t_j)\), which is equivalent to
  $\sum_{k=1}^d w_k \left( t_i[k] - t_j[k] \right) > 0.$
For distinct feature vectors,  equality defines a hyperplane through the origin;
its two   halfspaces determine which item ranks first. Each  ranking region is
  an intersection of comparison halfspaces with the positive unit ball, hence
convex. Intersecting it with a nonconvex acceptable region $\mathcal W^*$
may disconnect it.
Two full-dimensional ranking regions are \emph{adjacent} when their closures
share a facet. In Example~\ref{Example:2}, the neighboring regions are
adjacent in weight space, and their rankings differ only at lower positions.
A \emph{ranking distance}, defined below, measures how much the rankings differ.

\subsection{The  Exact  Stability of a Ranking}
The   \emph{exact stability}  of a ranking is the volume  fraction  of its ranking
region   within the acceptable weight region. Here $Vol$ denotes
$d$-dimensional Lebesgue volume. Equivalently, exact stability is the
probability that a uniformly chosen acceptable weight vector reproduces
the ranking exactly.

\begin{customdefinition}{2} [  Exact Stability in $\mathcal W^*$ ]
Let $\mathfrak{r}$ be a ranking on $\mathcal{D}$.   Its exact stability within
$\mathcal W^*$ is
 \begin{equation}
 Stab  _{\mathcal D,\mathcal W^*} (  \mathfrak r )
 =  \frac{Vol(\mathcal R^*_{\mathcal D}(\mathfrak r))}{Vol(\mathcal W^*)}.
 \end{equation}
\end{customdefinition}
 We write $Stab_{\mathcal D}(\mathfrak r)$ when
$\mathcal W^*=\mathcal W$.

 \subsection{The General Stability of a Ranking}

 The idea is straightforward: in evaluating the general stability of a ranking,
we do not just look at the volume of its own region in weight space. Rather,
we take into account the volumes of all rankings in the acceptable region
$\mathcal{W}^*$. Specifically, we compute the weighted sum of the volumes of
all ranking regions, where the weights decrease exponentially as the distances
between the ranking of interest and other rankings increase.

 Let $\mathfrak{R}_{\mathcal{D}}$ denote the set of all possible rankings generated by linear scoring functions on  $\mathcal{D}$.   A  distance function on $\mathfrak{R}_{\mathcal{D}}$   measures their dissimilarity: a larger  $dist(\mathfrak{r}_1, \mathfrak{r}_2)$ indicates   greater difference. We require two properties.

\begin{customdefinition}{3} [A distance function for rankings] \label{def:3}
A distance function for rankings on $\mathcal{D}$ is a function $dist: \mathfrak{R}_{\mathcal{D}} \times \mathfrak{R}_{\mathcal{D}} \to \mathbb{R} \cup \{+\infty\}$, which satisfies the following properties:
\begin{enumerate}
    \item For any $\mathfrak{r} \in \mathfrak{R}_{\mathcal{D}}$, $dist(\mathfrak{r}, \mathfrak{r}) = 0$
    \item For any $\mathfrak{r}_1, \mathfrak{r}_2 \in \mathfrak{R}_{\mathcal{D}}$, $dist(\mathfrak{r}_1, \mathfrak{r}_2) = dist(\mathfrak{r}_2, \mathfrak{r}_1) \geq 0$
\end{enumerate}
\end{customdefinition}

  We do not require the triangle inequality, and two distinct rankings may
have distance 0. Rankings retain item
identities, so a distance may distinguish which items exchange positions.
For target $\mathfrak r_0$, we weight each ranking region's volume by its ranking's \emph{credit},
$e^{-dist(\mathfrak r_0,\mathfrak r)}$: the target earns 1, and distant
rankings earn less.

Let $\mathfrak R^*_{\mathcal D}$ denote the rankings induced within
$\mathcal W^*$.
 \begin{customdefinition}{4} [  General Stability in $\mathcal W^*$ ]   \label{def:5}
For a target ranking $\mathfrak r_0\in\mathfrak R_{\mathcal D}$, its general
stability under distance  $dist$ is
 \begin{equation}
Stab  _{\mathcal D,dist,\mathcal W^*} (  \mathfrak r _0)
=  \frac{\sum_{\mathfrak r\in\mathfrak R^*_{\mathcal D}}
Vol(\mathcal R^*_{\mathcal D}(\mathfrak r))e^{-dist(\mathfrak r_0,\mathfrak r)}}{Vol(\mathcal W^*)}.
 \end{equation}
\end{customdefinition}
  We write $Stab_{\mathcal{D},dist}$ only when $\mathcal W^*$ is fixed by  context.

The   region volume fractions sum to  1,   and each credit lies in $[0,1]$.
Their weighted average is well-defined without the triangle inequality.
Equivalently, general stability is the expected credit of the ranking
induced by a uniformly chosen weight vector in $\mathcal W^*$.

 \begin{customexample}{3} \label{example:general_stability}

    Suppose we have four possible rankings  $\mathfrak{r}_0$, $\mathfrak{r}_1$, $\mathfrak{r}_2$, $\mathfrak{r}_3$ in an acceptable region $\mathcal{W}^*$ . The volumes of   their regions  are 2, 3, 3, and 2   in arbitrary units, totaling 10. Figure~ \ref{fig:2} presents the rankings and their distances as a graph.

\begin{figure}[H]
\centering
  \begin{tikzpicture}[
    scale=0.94,
    every node/.style={transform shape},
    ranknode/.style={circle, draw, fill=gray!12, inner sep=1pt, align=center},
    vtwo/.style={ranknode, minimum size=0.78cm},
    vthree/.style={ranknode, minimum size=1.12cm},
    edgelabel/.style={fill=white, inner sep=1pt}
]
\node[vtwo] (r0) at (0,0) {$\mathfrak{r}_0$\\[-1pt] {\footnotesize $V=2$}};
\node[vthree] (r1) at (2.05,0.78) {$\mathfrak{r}_1$\\[-1pt] {\footnotesize $V=3$}};
\node[vthree] (r2) at (2.05,-0.78) {$\mathfrak{r}_2$\\[-1pt] {\footnotesize $V=3$}};
\node[vtwo] (r3) at (5.95,0) {$\mathfrak{r}_3$\\[-1pt] {\footnotesize $V=2$}};
\draw (r0) -- node[edgelabel, above left] {\footnotesize 1} (r1);
\draw (r0) -- node[edgelabel, below left] {\footnotesize 1} (r2);
\draw (r1) -- node[edgelabel, right, xshift=12pt, yshift=-6pt] {\footnotesize 2} (r2);
\draw (r1) -- node[edgelabel, above right] {\footnotesize 4} (r3);
\draw (r2) -- node[edgelabel, below right] {\footnotesize 4} (r3);
\draw (r0) -- node[edgelabel, above] {\footnotesize 5} (r3);
\end{tikzpicture}
 \caption{Graph representation of the rankings and distances in Example ~ \ref{example:general_stability}.} \label{fig:2}
  \Description{A graph with four ranking nodes labeled r0, r1, r2, and r3. The nodes include volume labels, and edges between them are labeled with distances 1, 2, 4, and 5.}
 \end{figure}

\vspace{-10pt}
 Under exact stability,  both $\mathfrak{r}_0$ and $\mathfrak{r}_3$   have stability  $0.2$.

 Their general stability differs:
    \[
    Stab_{\mathcal{D}, dist}(\mathfrak{r}_0)
    =0.2+0.6e^{-1}+0.2e^{-5}=0.4221,
    \]
    while $Stab_{\mathcal{D}, dist}(\mathfrak{r}_3)
    =0.2+0.6e^{-4}+0.2e^{-5}=0.2123$. Both targets receive full credit
    from their own regions. The difference comes mainly from
    $\mathfrak r_1$  and   $\mathfrak r_2$, which together occupy 60\% of
    the acceptable region. They earn credit $e^{-1}$ relative to
    $\mathfrak r_0$, but only $e^{-4}$ relative to $\mathfrak r_3$.
    Thus more acceptable weights support rankings similar to
    $\mathfrak r_0$, even though the two targets have equal exact stability.
\end{customexample}

\subsubsection{ Exact  Stability is  a Special Case of General Stability}

   \begin{customdefinition}{5}[$0/\infty$ distance function]
  For rankings $\mathfrak{r}_1,\mathfrak{r}_2\in\mathfrak{R}_{\mathcal D}$, define
 \begin{equation*}
    dist_{0/\infty}(\mathfrak{r}_1, \mathfrak{r}_2)= 
    \begin{cases}
        0,    & \text{if } \mathfrak{r}_1 = \mathfrak{r}_2\\
        +\infty& \text{if } \mathfrak{r}_1 \neq \mathfrak{r}_2
    \end{cases}
\end{equation*}
\end{customdefinition}

  This function satisfies  Definition~\ref{def:3}. With this distance, general stability reduces to exact
stability~ \cite{asudeh2018obtaining}.

\subsection{The Algorithmic Problem }
We refer to the task of computing the stability value of a given ranking as \textit{stability computation}. This task is called \textit{stability verification} in prior work~\cite{asudeh2018obtaining}; we use the term computation to emphasize that the output is a stability   value  rather than a binary certificate.

  \begin{customproblem}{1}[general stability computation]
    Given a database $\mathcal D$  with $n$ items over $d$ scoring attributes,
    an acceptable region   $\mathcal W^*\subseteq\mathcal W$, an inducible ranking
    $\mathfrak r\in\mathfrak R_{\mathcal D}$, and a distance function $dist$,
    compute the general stability   $Stab_{\mathcal D,dist,\mathcal W^*}(\mathfrak r)$.
\end{customproblem}

\section{Practical Distance Functions} \label{sec:dist_example} \label{sec:dist_prop}

 The distance function must reflect which ranking changes matter in the application.
  When degrees of difference matter, it should offer  a spectrum
of  values rather than a binary distinction.   Discrepancies  at higher ranks   may
deserve greater weight:  shifts in the
top 10 universities   may be  more impactful than changes among those ranked
low.   We consider distances that
emphasize leading positions, accumulate disagreements, or preserve particular decisions.

  \paragraph{Top-$k$ distances.}
A user interested in the top 100 universities may care whether the same universities remain there, regardless of their order.
Prior work distinguishes two versions of top-$k$ stability: preserving the top-$k$ set or preserving the ordered top-$k$ list~\cite{asudeh2018obtaining}.
The \emph{top-$k$ set distance} is $0$ when two rankings have the same top-$k$ items and $\infty$ otherwise.
If their order also matters, the \emph{ordered top-$k$ distance} is $0$ when both the top-$k$ items and their order agree,  and   $\infty$ otherwise.
General stability under these distances recovers top-$k$-set stability and ordered top-$k$ stability, respectively: the probability that the corresponding set or ordered list stays unchanged.
These distances express a requirement to preserve that outcome exactly; other choices can give partial credit when appropriate.

 \subsection{Exponentially Position-Weighted Kendall's Tau and Spearman's Footrule Distances} \label{sec:pos_weight}

  We use  position-weighted versions of Kendall's Tau and Spearman's Footrule
  to make the same local change count less near the bottom.
For example, swapping the first two universities may matter more than
swapping those at ranks 99 and 100.

\subsubsection{Classical Kendall's Tau and Spearman's Footrule Distances} 

Kendall's Tau and Spearman's Footrule are established metrics for quantifying the dissimilarity between two rankings. Given two rankings $\mathfrak{r}_1, \mathfrak{r}_2$ over a set of items $\mathcal{D}$, the \emph{Kendall's Tau distance} counts the number of pairwise disagreements:
\[
\operatorname{dist}_{K}(\mathfrak{r}_1, \mathfrak{r}_2) = \sum_{\substack{(t_i, t_j) \in \mathcal{D} \times \mathcal{D} \\ \mathfrak{r}_1(t_i) < \mathfrak{r}_1(t_j)}} \mathbf{1}\left[ \mathfrak{r}_2(t_i) > \mathfrak{r}_2(t_j) \right],
\]
where $\mathbf{1}[ \cdot ]$ is the indicator function.

The \emph{Spearman's Footrule distance} measures the sum of absolute differences in positions:
  $\operatorname{dist}_{F}(\mathfrak{r}_1, \mathfrak{r}_2) = \sum_{t \in \mathcal{D}} \left| \mathfrak{r}_1(t) - \mathfrak{r}_2(t) \right|.$

\subsubsection{Position Weights}
  Both distances  treat discrepancies at all ranks equally. To emphasize discrepancies at higher ranks, \citet{kumar2010generalized} introduced \emph{position-weighted} versions of Kendall's Tau and Spearman's Footrule distances.   We use an
adaptation that assigns an unchanged item its local position weight,
whereas they assign it weight 1.

  \begin{customdefinition}{6}[Position-Weighted Kendall's Tau and Spearman's Footrule Distance]  \label{def:6} \label{def:weighted-kf}
 Let $\Delta = (\delta_1, \delta_2, \dotsc, \delta_n)$ be a sequence of non-increasing weights with $1 = \delta_1 \geq \delta_2 \geq \dotsb \geq \delta_n \geq 0$. Define the cumulative weight function $p(i) = \sum_{k=1}^i \delta_k$. For each item $t_i$, set
\[
\bar{p}(t_i) = \frac{p\big(\mathfrak{r}_1(t_i)\big) - p\big(\mathfrak{r}_2(t_i)\big)}{\mathfrak{r}_1(t_i) - \mathfrak{r}_2(t_i)}.
\]
When $\mathfrak{r}_1(t_i)=\mathfrak{r}_2(t_i)$, we use the convention $\bar{p}(t_i)=\delta_{\mathfrak{r}_1(t_i)}$.
Then, the \emph{position-weighted Kendall's Tau distance} is defined as
\[
\operatorname{dist}_{K_\Delta}(\mathfrak{r}_1, \mathfrak{r}_2) = \sum  _{\substack{(t_i, t_j) \in \mathcal{D}\times\mathcal D \\ \mathfrak{r}_1(t_i) < \mathfrak{r}_1(t_j)}}  \bar{p}(t_i) \bar{p}(t_j) \cdot \left[ \mathfrak{r}_2(t_i) > \mathfrak{r}_2(t_j) \right].
\]
The \emph{position-weighted Spearman's Footrule distance} is defined as
\[
\operatorname{dist}_{F_\Delta}(\mathfrak{r}_1, \mathfrak{r}_2) = \sum_{i=1}^n \bar{p}(t_i) \left| \sum_{j=1}^{\mathfrak{r}_1(t_i)} \bar{p}\big( \mathfrak{r}_1^{-1}(j) \big) - \sum_{j=1}^{\mathfrak{r}_2(t_i)} \bar{p}\big( \mathfrak{r}_2^{-1}(j) \big) \right|.
\]
  \end{customdefinition}

 We abbreviate these distances as $K_\Delta$ and $F_\Delta$.
For a moved item, $\bar p$ averages position weights along the ranks it crosses.
A stationary item can still reverse comparisons: $A>B>C$ and $C>B>A$ both
place $B$ second, while its comparisons with $A$ and $C$ reverse. We assign
it local weight $\delta_2$.

For example, with $\Delta=(1,1/2,1/4,1/8)$, swapping the first two items gives
both moved items weight $\bar p=p(2)-p(1)=1/2$, hence position-weighted Kendall and
Footrule distances $1/4$ and $1/2$. Swapping the last two gives both items weight
$1/8$ and distances $1/64$ and $1/32$. The same adjacent swap therefore matters
less near the bottom of the ranking.

 \subsubsection{Exponential Position Weights} 
A remaining issue is the choice of the weight sequence $\Delta$.
To  emphasize higher ranks   with a single parameter, we let each successive position receive a fixed fraction of the previous weight.
We use \emph{exponential position weights}:
  $\delta_k=\delta^{k-1}$ for $k=1,2,\dotsc,n$, where $0<\delta<1$.
We write  $K_\delta$ or $F_\delta$   for the resulting
 position-weighted Kendall   or Footrule distance. In particular, $K_{0.5}$   and $F_{0.5}$
set $\delta=0.5$, giving position weights $(1,0.5,0.25,\dotsc)$.
We use this setting in illustrations and also evaluate slower decay.
Values closer to 1 give appreciable weight to more positions.

  \subsection{Priority Distance}
\label{sec:priority-distance}

Rankings often determine decisions, but different rankings can often lead to the
same decisions. In such situations, we want stability to reflect whether the
decisions change and how important those changes are.
We propose \emph{priority distance}, denoted as $dist_{\rm pri}$, to express this idea.

Suppose the top ten applicants are admitted and the top three receive
scholarships. Two rankings may admit the same ten applicants and award
scholarships to the same three, despite ordering them differently.
Furthermore, suppose we regard a change in who is admitted more
important than a change in who receives a scholarship.

Priority distance assigns a nonnegative penalty, possibly $\infty$, to each
requirement to preserve a decision. It is zero if all requirements hold and
otherwise takes the largest penalty among the failed requirements.
In the example, assign penalty $\infty$ to changing the admitted group and
$\log 2$ to changing the scholarship recipients. Swapping ranks 1 and 2
preserves both decisions, giving distance 0. Swapping ranks 3 and 4 changes
only scholarships, giving distance $\log 2$. Swapping ranks 10 and 11 changes
admission, giving distance $\infty$.

Each requirement consists of fixed pairwise comparisons that must all hold.
For example, every admitted applicant must rank above every rejected applicant. Each pair inherits
the largest penalty of a requirement containing it, or 0 if none does.
Holding these pair penalties $a_{ij}$ fixed, define
\begin{equation}
 dist_{\rm pri}(\mathfrak r_1,\mathfrak r_2)
 =\max\bigl(\{a_{ij}:\mathfrak r_1,\mathfrak r_2
 \text{ disagree on }\{t_i,t_j\}\}\cup\{0\}\bigr).
 \label{eq:priority-distance}
\end{equation}
The fixed penalties make the  distance   symmetric and nonnegative;
requirements are specified for the target ranking before evaluating its stability.

\subsection{First Differing Position Distance}
\label{sec:dist_pos}

 The   \emph{first differing position distance}, denoted as $dist_{pos}$,
assigns smaller distances to rankings that agree on more top-ranked items.
It is zero for identical rankings and $1/k$ otherwise, where $k$ is the
first position at which they differ:
\begin{equation}
 dist_{pos}(\mathfrak r_1,\mathfrak r_2)=
 \begin{cases}
 0,&\mathfrak r_1=\mathfrak r_2,\\
 1/k,&\text{otherwise.}
 \end{cases}
 \label{eq:dist-pos}
\end{equation}
For example, $(a,b,c,d,e,f)$ and $(a,b,c,f,d,e)$ first differ at position 4,
so their distance is $1/4$.
This distance emphasizes agreement at the top and ignores later disagreements.
It never exceeds 1, regardless of database size or dimension, and its
distance thresholds define convex weight regions. Section~\ref{sec:structured-computation}
explains how these two properties support efficient computation.

\section{2D Stability Computation}
 \label{sec:2d}

Like \cite{asudeh2018obtaining}, we first consider   $d=2$.
A weight direction is  represented by one angle.   The angles at which two items
exchange order  partition the angular   range  into intervals,   each inducing one ranking.
For general stability, we sweep these intervals and   weight each angular width
by the induced ranking's credit.

  Let $\theta_0<\cdots<\theta_q$ be the distinct exchange angles and endpoints
of the acceptable range, and let $\mathfrak s_j$ be the ranking on
$(\theta_j,\theta_{j+1})$. The sweep $GSC_{2d}$ returns
\[
 \frac{\sum_{j=0}^{q-1}(\theta_{j+1}-\theta_j)
 e^{-dist(\mathfrak r_0,\mathfrak s_j)}}{\theta_q-\theta_0}.
\] 
  This is exact for the full positive disk or a fixed-radius radial sector:
ranking is constant along each ray, and normalized area equals normalized
angular width. Other regions require direction-dependent radial weighting.

Each pair exchanges order at most once, giving $O(n^2)$ intervals. Sorting these angles
and updating groups of tied items takes $O(n^2\log n)$ time,   and each interval requires
one distance evaluation. Thus the total is $O(n^2(D_{dist}+\log n))$, with
$O(n^2)$ extra space, where $D_{dist}$ is the distance-evaluation cost.
Exactness assumes exact tests for ordering the exchange angles and exact interval integration.
The artifact gives the implementation, including how simultaneous exchanges are handled.

\section{Multidimensional Stability Computation}
 \label{sec:multi}

  Computing stability exactly  is $\#P$-hard   when dimension is part of the input.
We therefore extend the sampling method of \citet{asudeh2018obtaining}:
the resulting algorithm, $GSC_{md}$, samples acceptable weights and evaluates their rankings' credits
until it can estimate general stability at the requested accuracy.

  \paragraph{Hardness of exact computation.}
It suffices to consider exact stability on the full positive ball. For a
partial order $P$ on $d$ coordinates with constraints $(a_h,b_h)$,
$1\leq h\leq m$, construct items
$v_\ell=\sum_{h=\ell}^m(e_{a_h}-e_{b_h})$ and $v_{m+1}=0$, where $e_j$
is coordinate vector $j$. Away from ties, the target
$v_1\succ\cdots\succ v_{m+1}$ holds exactly when all $w_{a_h}>w_{b_h}$.
Coordinate symmetry gives stability $e(P)/d!$, where $e(P)$ counts
linear extensions (total orders respecting $P$), a $\#P$-complete counting problem~\cite{brightwell1991counting}.
A strict linear extension induces the target. A common translation
and positive scalar rescaling place the items in $[0,1]^d$ with
polynomial encoding length, so exact stability recovers $e(P)$ even for
these rational inputs.

  \subsection{Sampling Acceptable Weights}

  We sample weights uniformly from $\mathcal W^*$ so that a ranking appears
 with probability equal to   its  region's volume fraction. For the full positive
ball, we use the sampler of \citet{asudeh2018obtaining}: draw $d$ independent
standard Gaussian coordinates, take their absolute values, and normalize
the vector to unit length. The resulting directions give the same ranking
probabilities as uniform volume, because positive rescaling preserves rankings.
For angular restrictions, \citet{asudeh2018obtaining} also give conditional
sampling methods. For a general $\mathcal W^*$,  $GSC_{md}$   takes an independent
uniform-volume sampler as input.

  \subsection{\texorpdfstring{$GSC_{md}$}{GSCmd}: A General Baseline}
\label{sec:adaptive-interval}

 $GSC_{md}$   repeatedly draws an acceptable weight vector, ranks the items,
and computes the resulting ranking's credit. Averaging these credits
estimates general stability. For a uniformly sampled weight vector $W\in\mathcal W^*$, write
$X=\exp[-dist(\mathfrak r_0,\nabla_{f_W}(\mathcal D))].$
After $N\geq1$ independent samples, the estimate is
$\widehat\mu=N^{-1}\sum_{i=1}^N X_i$.
Each ranking occurs with probability equal to its region's volume fraction,
so $E[\widehat\mu]=\mu$, the   requested general stability.

  More samples improve average accuracy but require more computation.
Let $C_{\mathcal W}$ be the cost of generating one weight vector and $C_{draw}(n,d)$
the total cost per sample, including weight generation, scoring, ranking, and distance evaluation. Then  $N$   samples cost
$O(NC_{draw}(n,d))$.

  The \emph{relative credit variance} $H_{dist}$ is credit variance divided
by its squared mean. Larger values need more samples for equal accuracy.

 For  fixed   $N$, independence gives $Var(\widehat\mu)=Var(X)/N$.
Thus, for $\mu>0$, the relative mean-square error is
\begin{equation}
 \frac{E[(\widehat\mu-\mu)^2]}{\mu^2}
 =\frac{H_{dist}}{N},\qquad
 H_{dist}:=\frac{Var(X)}{E[X]^2}.
 \label{eq:relative-credit-variance}
\end{equation} 
  \paragraph{Stopping at the requested accuracy.}
The user specifies the tolerated relative error  $\epsilon$ and   confidence
$1-\alpha$. Algorithm~\ref{alg:direct-gsc} draws weights in batches and updates
a confidence interval for stability. It stops when the interval is narrow
enough to return an estimate within $(1\pm\epsilon)\mu$, for $\mu>0$.
We state the interval rule for independent, identically distributed (iid) sample values $Y\in[a,b]$;
 $GSC_{md}$   uses $Y=X$. Let $\overline Y_N$  and   $v_N$ be their sample mean
and unbiased sample variance (divisor $N-1$).
Use $[a,b]=[0,1]$ unless a tighter credit range is known.
At predetermined, geometrically increasing checkpoints $N_j\geq2$, $j=0,1,\ldots$,
allocate $\alpha_j=6\alpha/[\pi^2(j+1)^2]$,   $0<\alpha<1$, and use~\cite{maurer2009empirical}
\[
 h_j=\sqrt{\frac{2v_{N_j}\log(4/\alpha_j)}{N_j}}
       +\frac{7(b-a)\log(4/\alpha_j)}{3(N_j-1)}.
\]
Clip $[\overline Y_{N_j}-h_j,\overline Y_{N_j}+h_j]$ to $[a,b]$.
Applying the bound to both tails and summing $\alpha_j$ gives
$1-\alpha$ coverage at all checkpoints.
Stop at the first confidence interval $[L,U]$ for stability with $0<L\leq U$ and
$r=(U-L)/(U+L)\leq\epsilon$ and return $2LU/(L+U)$, whose worst-case
relative error is $r$ at both endpoints.
Report incomplete if the budget runs out first.
For $\mu=0$, relative error is undefined; zero observations alone do not
certify zero stability.

  \begin{algorithm}[bp]
 \caption{$GSC_{md}$   at requested accuracy }
  \label{alg:direct-gsc}
\KwData{$\mathcal D$, target $\mathfrak r_0$, $dist$, iid sampler for $\mathcal W^*$,
$\epsilon$, $\alpha$, credit range $[a,b]$, checkpoints $N_j$}
\For{each checkpoint $N_j$ allowed by the sample budget}{
  draw fresh acceptable weights up to a total of $N_j$ samples;
  compute each induced ranking's credit $X$\;
  update the sample mean and variance\;
  form $[L,U]$ with confidence radius $h_j$, clipped to $[a,b]$\;
  \If{$0<L\leq U$ and $(U-L)/(U+L)\leq\epsilon$}{
    \KwRet $2LU/(L+U)$ and $[L,U]$\;
  }
}
 \KwRet   incomplete \;
\end{algorithm}

  \subsection{Exponential Worst-Case Sampling Cost}
 \label{sec:not_poly}

  Unfortunately,  $GSC_{md}$   can require exponentially many samples. For exact
stability, $X=1$ if the sampled ranking matches the target and $X=0$ otherwise.
A target region of volume fraction
$p$ needs $\Omega(1/p)$ samples for fixed relative error: fewer samples have
constant probability of no match.

  \begin{customproposition}{1}
\label{prop:direct-hard}
For fixed $d$, there are  $n$  -item databases  whose positive weight   orthant
contains $\Omega(n^{2(d-1)})$  nonempty ranking regions.  Consequently, on some
exact-stability query, $GSC_{md}$ requires $\Omega(n^{2(d-1)})$ uniform weight samples for fixed
relative error and confidence. For $d=n$, the items $e_1,\ldots,e_n$ induce
$n!$ equal-volume rankings, so $GSC_{md}$ requires $\Omega(n!)$ samples for each exact target.
 \end{customproposition}

\begin{proof}
 Cover's   count~\cite{cover1967linearly} gives generic $n$-point inputs in
$\mathbb R^d$ with $\Theta(n^{2(d-1)})$ induced orderings for fixed $d$.
Some coordinate orthant contains a $2^{-d}$ fraction of these regions;
sign flips map it to the positive orthant. One region has
volume fraction $O(n^{-2(d-1)})$, and the probability of observing no match gives the
sample lower bound. A common translation and positive scalar rescaling put
attributes in $[0,1]$ without changing rankings.
For $e_1,\ldots,e_n$, item $i$ has score $w_i$; coordinate symmetry gives
every order probability $1/n!$.
  \end{proof}

\section{Efficient Computation of General Stability}
\label{sec:structured-computation}

  Section~\ref{sec:multi} introduces $GSC_{md}$, which applies to any distance function but can take exponential time in the worst case.
We now examine three distance properties that allow faster computation.
First, we find that $GSC_{md}$ itself can be efficient for distances with a property we call \emph{uniform boundedness}.
For distances with a second property, \emph{tail decay with bounded influence}, we introduce Prefix-GSC, an optimization that reduces sorting per sample.
For distances with a third property, which we term \emph{quasiconvexity}, we propose Conv-SC, which computes general stability by approximating convex volumes.
Table~\ref{tab:method-map} summarizes the properties, their computational benefits, and example distances.

  \begin{table*}[tbp]
\centering
\small
\textbf{Choosing a method for general stability computation}\par\smallskip
\setlength{\tabcolsep}{3.2pt}
\begin{tabular}{@{}>{\raggedright\arraybackslash}p{0.12\textwidth}>{\raggedright\arraybackslash}p{0.19\textwidth}>{\raggedright\arraybackslash}p{0.24\textwidth}>{\raggedright\arraybackslash}p{0.09\textwidth}>{\raggedright\arraybackslash}p{0.31\textwidth}@{}}
\toprule
\textbf{Case or property} & \textbf{Insight} & \textbf{Example distance} &
\textbf{Method} & \textbf{Computational cost or saving}\\
\midrule
\textbf{General case} &
& Classical Kendall and Footrule; $dist_{0/\infty}$ (rare target ranking) &  $GSC_{md}$   &
\textbf{Exponential worst-case sample count}
(Proposition~\ref{prop:direct-hard}).\\
\midrule
\textbf{Uniform boundedness} & Each sample contributes at least a fixed positive amount.
& $K_\delta,F_\delta$ (fixed $0<\delta<1$); $dist_{pos}$; $dist_{\rm pri}$ with a fixed finite penalty bound & $GSC_{md}$ &
\textbf{Sample budget independent of $n,d$}. Polynomial total time
with polynomial work per sample.\\
\midrule

  \textbf{Tail decay with bounded influence} & Sort leading items and bound the effect of the rest.
& $K_\delta,F_\delta$ (exponential weights) &  $GSC_{md}$   optimization:\newline Prefix-GSC &
\textbf{Avoid full sorting:} $O(n)$ rather than $O(n\log n)$ ranking work
for a fixed number of leading items.\\
\midrule

  \textbf{Quasiconvexity} & Volumes of convex bodies can be approximated efficiently.
& $dist_{0/\infty}$; top-$k$ distances; $dist_{\rm pri}$ (finite or infinite penalties); $dist_{pos}$ &  Conv-SC   &
\textbf{Polynomial time} under the volume-approximation conditions of
Theorem~\ref{thm:convsc-guarantee}.\\

  \bottomrule
\end{tabular}

  \caption{When general stability can be computed faster.
$dist_{pos}$ satisfies both uniform boundedness and quasiconvexity, so $GSC_{md}$ and Conv-SC can both estimate general stability in polynomial time under the stated conditions.}
\label{tab:method-map}
\label{tab:condition-overlap}
\end{table*}

\subsection{Uniform Boundedness: Efficient Sampling}
\label{sec:fewer-draws}

Section~\ref{sec:not_poly} establishes an exponential worst case for $GSC_{md}$, but its runtime need not be exponential for every distance.
We therefore look for distance properties that guarantee efficient computation.
For exact stability, $GSC_{md}$ may have to wait for rare matches, with most samples contributing zero.
If every sample instead contributes at least a fixed positive amount, we can bound the number needed to reach the requested accuracy.

We call a distance \emph{uniformly bounded} if
it never exceeds a constant $B$ independent of database size and dimension.
Every sampled ranking then contributes at least $e^{-B}$ to the general stability estimate.
Therefore, a fixed sample budget suffices as database size and dimension grow, for fixed bound, accuracy, and confidence.
For example, $K_{0.5}$ and $F_{0.5}$ give at least $e^{-9/8}\approx0.32$ per sample, as established below.

\begin{customthm}{2}[Sampling under uniform boundedness]
\label{thm:bounded-credit}
If $0\leq dist\leq B$, Algorithm~\ref{alg:direct-gsc} with credit range
$[e^{-B},1]$ and a fixed geometric checkpoint schedule meets its stopping
condition within a sample budget depending only on $B,\epsilon,\alpha$.
\end{customthm}
\begin{proof}
Put $c=e^{-B}$. The relative interval radius is at most $h_j/c$, and
  $v_N\leq N(1-c)^2/[4(N-1)]$.
Substitution into the confidence bound gives a radius tending to zero at
rate $O(\sqrt{\log(j+1)/N_j})$ for fixed $B,\alpha$.
Thus a qualifying checkpoint is independent of $n,d$.
\end{proof}

\paragraph{Bounding position-weighted Kendall and Footrule.}
For $n\geq2$ and $W_\Delta=\sum_{j=1}^n\delta_j$, every nonnegative,
nonincreasing, \emph{discretely convex} position-weight sequence
($\delta_j-2\delta_{j+1}+\delta_{j+2}\geq0$) bounds the position-weighted distances by
\begin{equation}
 K_\Delta,F_\Delta\leq
 B_\Delta:=\tfrac12(W_\Delta-\delta_1+\delta_2)^2.
 \label{eq:weighted-diameter}
\end{equation}
For any fixed exponential decay $\delta<1$, $W_\Delta<(1-\delta)^{-1}$,
so the bound is independent of $n,d$. Its constant grows as $\delta$
approaches 1. At $\delta=0.5$, $B_\Delta\leq9/8$.
The work per sample remains $O(C_{\mathcal W}+nd+n\log n)$.

\begin{proof}[Proof sketch]
Discrete convexity bounds an item's weight by the average of the position weights
at its two ranks. Remove the identical leading items, which contribute zero.
At the first differing rank $k$, the two items occupying that rank tighten the
bound on their combined weight by $\delta_k-\delta_{k+1}$, leaving total weight
$S\leq\sum_{j=k+1}^n\delta_j+\delta_{k+1}\leq W_\Delta-\delta_1+\delta_2$.
$K_\Delta$ is at most $S^2/2$.
For $F_\Delta$, pack each item's weight as an interval in each ranking on
$[0,S]$. Matching corresponding intervals gives
$F_\Delta\leq2\int_0^S|x-S/2|\,dx=S^2/2$.
\end{proof}

\subsection{Tail Decay with Bounded Influence}
\label{sec:cheaper-samples}
\label{sec:prefix-gsc}

In many ranking applications, changes near the top matter more than changes near the bottom.
Giving larger weights to leading positions expresses this preference, but fully sorting lower-ranked items spends time on differences that contribute little to the answer.
Prefix-GSC saves this work by sorting only leading items and bounding the effect of the others.
It applies when lower-ranked items have diminishing total weight and that weight has bounded influence on the distance.
All items are still scored; the saving is in constructing the ranking.

\paragraph{When it applies.}
Let $q_i=\bar p(t_i)$ be the item weights from Definition~\ref{def:weighted-kf}, and let $W$ bound the total position weight independently of $n,d$.
For fixed target ranks $r_i$ and sampled ranks $b_i$, write $D=D(q)$ for the weighted distance.
\emph{Bounded influence} means that reducing item weights by a total $J$ can only decrease $D$, by at most $LJ$, whenever $\sum_iq_i\leq W$.
Here $L$ is known and independent of $n,d$.
The distance must be evaluable from  the   nonzero-weight items and their full ranks in both rankings.
The \emph{tail-decay property} requires $T_m=\sum_{j>m}\delta_j$ to tend to zero uniformly over database size.

\paragraph{Procedure.}
For each sampled weight vector, score all items, then select and sort the sampled top $m$.
Retain the union of the target and sampled top-$m$ items.
The sampled top-$m$ ranks are now known.
To find the sampled ranks of the target top $m$, first sort these $m$ items by their sampled scores.
For each database item, use binary search to find its insertion position in this sorted list.
Count how many items have each insertion position, then use cumulative counts to recover the full sampled ranks of the target top-$m$ items.
A target top-ten item dropping to rank 40 is therefore treated at its full rank.
Compute retained item weights from these ranks and set other weights to zero, giving $q'$.
Evaluate $D_m=D(q')$.
One sample costs $O(C_{\mathcal W}+nd+n\log(m+1))$, plus distance evaluation on $\leq2m$ items.

\begin{customthm}{3}[Prefix-GSC error bound]
\label{thm:prefix-certificate}
For nonincreasing, discretely convex position weights and a distance with bounded influence $L$, Prefix-GSC gives
$0\leq D-D_m\leq B_m:=LT_m$.
Consequently, $e^{-B_m}e^{-D_m}\leq e^{-D}\leq e^{-D_m}$.
\end{customthm}

\begin{proof}
Convexity and monotonicity of the position weights give
$q_i\leq(\delta_{r_i}+\delta_{b_i})/2$.
Thus total item weight is at most $W$, and omitted weight is at most $T_m$, since both ranks of an omitted item exceed $m$.
Apply bounded influence to $q'$.
\end{proof}

\paragraph{Examples and tighter bounds.}
Write $W_n=\sum_{j=1}^n\delta_j$.
$K_\Delta$ satisfies bounded influence with $L=W$,  and   $F_\Delta$ with $L=2W$.
For $K_\Delta$, each item's marginal influence is at most the total weight of the other items.
For $F_\Delta$, index items by target rank and let $s_i$ be the sign of the difference between their target and sampled cumulative weights.
Within each fixed-sign region, $D(q)=\sum_{i<j:b_i>b_j}(s_j-s_i)q_iq_j$ with $0\leq s_j-s_i\leq2$; continuity gives $L=2W$.
Integrating these marginal bounds gives the sharper bound $B_m=W_nT_m-T_m^2/2$ for $K_\Delta$ and twice this for $F_\Delta$, which we use in the experiments.
Summing weights of items that change position also qualifies, with $L=1$ and $B_m=T_m$.

  \paragraph{Accuracy and stopping.}
Before sampling, choose $m\leq n$ so $B_m$ uses at most a fixed fraction of $\log((1+\epsilon)/(1-\epsilon))$, leaving room for sampling error.
If $[L_m,U_m]$ is a checkpoint confidence interval for the mean of $e^{-D_m}$,
then $[e^{-B_m}L_m,U_m]$ covers stability under the full distance.
Clip it to the known stability range and apply Algorithm~\ref{alg:direct-gsc}'s stopping rule.
This accounts for both sampling uncertainty and the omitted contribution.

Exponential position weights and power weights $\delta_j=j^{-p}$ with $p>1$ satisfy the tail-decay property.
At fixed distance parameters and accuracy, $m$ is independent of $n,d$.
For the examples above, the ranking work becomes $O(n)$; the shared $O(nd)$ scoring cost limits high-dimensional gains.

  \subsection{Quasiconvexity: Conv-SC}
\label{sec:convex}

The volume formulation of stability suggests another route beyond uniform
sampling. Although exact volume computation is \#P-hard, randomized geometric methods~\cite{smith1996hit}
can approximate convex-body volumes in polynomial time~\cite{dyer1991random,kannan1997random,lovasz2006simulated}.
We propose Conv-SC to estimate stability through these volume approximations.

Reproducing a target ranking requires every item to stay ahead of those
below it. Each comparison is a linear inequality in the scoring weights.
Their intersection within a convex acceptable region is convex; its volume
fraction gives exact stability.
Top-$k$ set distance requires each target top-$k$ item to stay ahead of every item outside that set.
Ordered top-$k$ distance additionally fixes the order within the top $k$.
Both therefore give convex regions by the same intersection argument.

General stability involves many ranking regions. Approximating each volume
separately could still require exponentially many calls. Conv-SC instead
groups  all ranking regions   within  distance $r$   of the target and estimates
their combined volume. Each combined region must
be convex: the \emph{quasiconvexity} requirement. Convex ranking regions can
have nonconvex unions; the position-weighted Kendall and Footrule distances can fail this condition.

  The admissions example (Section~\ref{sec:priority-distance}) needs two volume
fractions: the probability of preserving admission and that of preserving both
decisions. General stability is
$\tfrac12 P(\text{admission preserved})+
\tfrac12 P(\text{both preserved})$:
full credit for preserving both decisions, half for preserving only admission.
Both regions are convex intersections of the required comparison halfspaces.

  \begin{customdefinition}{7}[Quasiconvex distance]
For a database $\mathcal D$, a ranking distance is \emph{quasiconvex} if,  for
  every inducible target $\mathfrak r_0$ and radius  $r$, the
distance-threshold region in weight space is convex:
\[
B_{dist}(\mathfrak r_0,r)=
\{w\in\mathcal W:dist(\mathfrak r_0,\nabla_{f_w}(\mathcal D))\leq r\}.
\] 
  \end{customdefinition}

  First  differing position distance  is also quasiconvex: a threshold preserves
a fixed target prefix, an intersection of comparison halfspaces.
Its at most $n$ finite distance values admit  Conv-SC, while its bound of 1 also permits
efficient sampling.

  \begin{customproposition}{4}
\label{prop:priority-convex}
Priority distance $dist_{\rm pri}$ is quasiconvex on a convex acceptable region.
\end{customproposition}
 \begin{proof}
  A distance threshold $r$ preserves every target comparison with penalty
$a_{ij}>r$, intersecting open or closed comparison halfspaces with the
acceptable region. Identical items retain their fixed order; closing
other tie boundaries preserves volume.
 \end{proof}

  Keeping one named item somewhere in the top $k$ may instead form a
nonconvex union.

  \paragraph{Combining volumes into general stability.}
List the finite distance values as $0=r_0<\cdots<r_L$ and put $\xi_i=e^{-r_i}$,
$\xi_{L+1}=0$, and $C_i=B_{dist}(\mathfrak r_0,r_i)\cap\mathcal W^*$.
A ranking at distance $r_j$ belongs to $C_j,\ldots,C_L$; its coefficients
telescope to $\xi_j$. Thus
\begin{equation}
 \mu=\frac{\sum_{i=0}^{L}(\xi_i-\xi_{i+1})Vol(C_i)}{Vol(\mathcal W^*)}.
 \label{eq:convsc-cumulative}
\end{equation}
Rankings at infinite distance contribute zero. A max-margin program supplies
an interior ball for comparison bodies~\cite{boyd2004convex}.

Conv-SC   uses volume estimates $\widehat V(C;\eta,\nu)$ with relative error
$\eta$ and failure probability $\nu$, including geometric preparation.

  Construct each region $C_i$, set zero-volume terms to zero, and substitute
the remaining volume estimates into Equation~\ref{eq:convsc-cumulative}.
Estimate the denominator too if it is unknown; a zero estimate reports failure.

  \begin{customthm}{5}[Conv-SC guarantee]
\label{thm:convsc-guarantee}
Suppose $\mu>0$  and   $0<\epsilon\leq1$, with polynomially many explicitly described convex regions $C_i$. Their rational body
descriptions, that of convex $\mathcal W^*$, and nonnegative rational coefficients
have total bit length $M$.
Assume polynomial-time membership, separation, zero-volume recognition, and
affine rounding between concentric balls with polynomial radius ratio for
full-dimensional bodies. Let $T_V(d,M,\eta,\nu)$ be the cost of
$\widehat V(C;\eta,\nu)$.  Conv-SC returns relative error $\epsilon$ with failure probability
$\alpha$ in $O((L+2)T_V(d,M,\epsilon/4,\alpha/(L+2)))$ time plus
polynomial processing. Standard volume oracles make this a fully
polynomial randomized approximation scheme (FPRAS) under these conditions.
\end{customthm}

  \begin{proof}
With probability at least $1-\alpha$, all at most $L+2$ estimates are accurate
(union bound). Nonnegative coefficients preserve the numerator's relative
error $\eta$. A known denominator therefore permits $\eta=\epsilon$.
Otherwise division gives ratios between $(1-\eta)/(1+\eta)$ and
$(1+\eta)/(1-\eta)$; $\eta=\epsilon/4$ suffices.
\end{proof}

For $dist_{pos}$, each positive coefficient is at least $e^{-1}/n^2$.
Approximating it to $\epsilon/4$ relative error uses $O(\log(n/\epsilon))$ bits.
Applying Theorem~\ref{thm:convsc-guarantee} at tolerance $\epsilon/2$
bounds the combined relative error by $\epsilon$.

\paragraph{Computing the volumes in practice.}
Our  Conv-SC   implementation encloses each convex region in a box and samples inside
that box, concentrating effort on weights that can contribute to the answer.
Rescale positive-ball directions to $\sum_jw_j=1$ and construct an affine
enclosing box for the comparison polytope using linear programs.
Timings charge all candidate box constructions; the artifact gives their details.

Uniform box samples require a density correction.
On $\sum_jw_j=1$, the original directions have density
$f(w)=\frac{2^{d-1}\Gamma(d/2)}{\pi^{d/2}\|w\|^d}$ in coordinates $w_1,\ldots,w_{d-1}$.
For the region $C$, minimizing an affine lower bound on $\|w\|^2$ by
certified linear programming gives an upper density bound $M$.
Let $V$ be the box volume in these coordinates and $P(C)$ the region's
probability under the original weights.
A uniform sample from the box gives
\begin{equation}
 Y=\mathbf1[w\in C]f(w)/M\in[0,1],\qquad
 P(C)=\Lambda E[Y],\quad\Lambda=VM.
 \label{eq:conv-enclosure}
\end{equation} 
  Section~\ref{sec:adaptive-interval}'s interval for $E[Y]$, scaled by $\Lambda$,
controls the probability error. A fixed budget of
$O((\Lambda/P(C))\epsilon^{-2}\log(1/\alpha))$ samples suffices.
A tight box can give $\Lambda\ll1$, reducing the sampling needed for a rare
region; for $\Lambda\geq1$, we use direct weight sampling.
Split the failure budget $\alpha$ among the estimated regions and combine their
probability intervals through Equation~\ref{eq:convsc-cumulative}; uncomputed
probabilities remain in $[0,1]$ until the overall interval is sufficiently narrow.

  \paragraph{Sharing samples across regions.}
For priority queries with credits $1,1/2,0$, let $O$ preserve the higher-priority
requirements and $I\subseteq O$ preserve both. One box for $O$, using
$Y(1+\mathbf1[w\in I])/2$, estimates $\mu=[P(O)+P(I)]/2$ with the same confidence rule.
This saves a second box and separate estimation of a possibly tiny component.
Gains depend on how tightly the box encloses the region and the cost of constructing it. This implementation
can still take exponential time; Theorem~\ref{thm:convsc-guarantee}'s polynomial
guarantee uses its stated volume oracles.

  \subsection{Choosing a Method}
\label{sec:method-choice}
A distance can satisfy more than one property in Table~\ref{tab:method-map}, so more than one method may apply.

 For $K_{0.5}$ and $F_{0.5}$, start with $GSC_{md}$ using the distance bound and consider
Prefix-GSC for large databases. Both $GSC_{md}$ and Conv-SC apply to $dist_{pos}$;
the measured comparison favors $GSC_{md}$
(Section~\ref{subsec:direct-experiment}). For top-$k$ distances, Conv-SC is an option when the unchanged top-$k$ outcome is rare.
When several methods qualify, compare total
time at the same requested accuracy, including setup. Use fresh samples
after any preliminary method selection.

\section{Experiments}
\label{sec:experiments}

  Our experiments examine both the usefulness of general stability and the cost of computing it.
We first use real rankings to show what general stability reveals about different changes  that exact stability   treats alike.
We then test the accuracy of  the approximation methods   and compare their runtimes at the same requested accuracy.
The runtime experiments examine when $GSC_{md}$ is efficient, when Prefix-GSC saves additional work, and when  Conv-SC   benefits from computing convex volumes.

\subsection{ Setup }
\label{subsec:exp_1}

  Runs use  an Intel i7-12700F, 16 GB   RAM, and Python 3.11/ C++.
University timing studies
use up to 500 items from QS 2023 (six attributes)~\cite{qsranking},
the  Center for World University Rankings (CWUR) 2015  (eight) ~\cite{cwur2015},
  and Times Higher Education (THE)  2016   (five)~\cite{the2016}.
The  Multidimensional Poverty Index (MPI)   2023 scaling study uses 30,  50, and 100 rows with up to ten
attributes~\cite{un_mpi}. These four sources and
 CSRankings 2024~\cite{csrankings2024}, NBA 2023--24
~\cite{basketballreference2024}, and the Environmental Performance Index (EPI) 2024~\cite{epi2024}   form the seven-source
collection.\footnote{CWUR component ranks become $1/(1+\mathrm{rank})$.
QS and EPI require named rows with at least four and three scores,
respectively, then zero-fill missing fields. MPI rows are country--survey
records from UNDP's Table~2, with missing indicators zero-filled.}

By default, real-data queries use the first $n$ items remaining after preprocessing, in source order
(MPI: sorted by a fixed weighted sum; NBA: first regular-season row per player), then min--max scale
attributes to $[0,1]$ (constants become zero). Queries with nonfinite attribute values are excluded without substituting other inputs.
Targets use equal weights; directions are uniform on the positive orthant of
the unit sphere;
generated coordinates are iid uniform $[0,1]$. Stored row indices serve as item IDs.\looseness=-1\relax

Unless stated otherwise, algorithm accuracy and runtime tests request 2\% relative error at 95\% confidence
using Section~\ref{sec:adaptive-interval}'s stopping rule. The artifact
specifies each run's checkpoints and sample cap.
The scaling study marks time-capped runs; all other reported adaptive runs
finish at the requested accuracy. Comparisons of Prefix-GSC with $GSC_{md}$ include setup,
loading, worker startup, confidence intervals, and sampling.
Shared preprocessing, target construction, and interpreter startup precede these clocks.
Scaling timings exclude common library initialization but include query-specific geometry and bounds.
Percentile bootstraps of paired time ratios give 95\% intervals for median speedup.
The GitHub artifact provides inputs, attribute order, row-filtering rules,
query settings, code, and results.

\subsection{  What General Stability Reveals }
\label{subsec:real_world_cases}

  We first examine how the choice of distance changes general stability, then use examples to show when these differences are useful.
Usefulness depends on which ranking changes matter in an application; no single stability measure is best for every user.

\paragraph{Does the distance change which rankings are more stable?}
Exact stability is a special case of general stability: its $0/\infty$ distance gives every change, however small, an infinite penalty.
We ask whether using $dist_{pos}$,  $K_{0.5}$, or $F_{0.5}$ instead changes which target rankings are considered more stable.
These distances give less importance to changes near the bottom of a ranking.

For each of the seven datasets, we use database sizes 20  and   30 with the first three attributes.
This small, three-dimensional setting lets us compute stability by integrating all ranking regions, including those with very small volumes.
For each dataset and size, 30 random positive weight vectors generate the target rankings, giving 420 queries.
For every target,  we compute four stability values, one under each distance, over the same positive weight space.
We pool these queries and compute the Spearman rank correlation of exact stability with each of the other three stability values.
A correlation near 1 means that the distance choices largely agree about which targets are more stable.

The correlations with exact stability are only 0.276 for  $dist_{pos}$, 0.252 for  $K_{0.5}$, and 0.256 for  $F_{0.5}$ .
Thus the choice of distance substantially changes which rankings are assessed as more stable.
Within the 14 dataset--size groups, the median correlations are also low: $-0.013$ for $K_{0.5}$ and $-0.024$ for $F_{0.5}$.
Examples below explain when that difference is useful.

  \paragraph{A ranking with low exact stability can still identify the same leading universities.}
Consider the  30 universities with the highest   Patents scores in CWUR, ranked using equal weights on eight attributes.\footnote{We scale all eight attributes on the 1,000 finite rows before selecting the Patents top 30, then rescale this subset.}
 The target begins with Harvard, MIT, Stanford, Johns Hopkins, and UC Berkeley.
  We draw two million independent weights over the full positive weight space and compare each induced ranking with this target.
Only ten samples reproduce the full ranking, giving estimated exact stability $5.0\times10^{-6}$.
Yet general stability under $K_{0.5}$ is 0.9933, suggesting little disagreement when leading positions matter most.

  To understand this difference, we inspect what happens to the five leading universities.
All five remain in the top five in 69.30\% of samples, and at least four remain in 98.00\%.
The average fraction remaining is 93.46\%.
The ranking is unstable under exact stability, but nearly the same universities remain at the top.
For a user interested in identifying leading universities, general stability captures the agreement on the top positions.

  \paragraph{A ranking with low top-$k$-set stability can still preserve most top-$k$ items.}
Top-$k$-set stability is itself general stability under top-$k$ set distance.
This choice captures part of the agreement in the CWUR example, but it treats even one replacement as a complete mismatch.
When a few replacements are acceptable, this distance may also be too strict.
We examine 500 QS universities and small changes to the equal target weights.
We draw $2^{20}$ weights uniformly from the positive part of a Euclidean ball of radius $0.1\|w_0\|$ around the target vector $w_0$, where $\|w_0\|=1/2$.
Only 7.05\% of samples keep exactly the same top 100 universities, but an average of 98.36 of the original 100 remain.
The top-100 set changes frequently, yet very few universities are replaced on average.

  General stability can express the extent of these changes.
Let $q$ be the fraction of the target's top-$k$ items still in the top $k$; keeping 98 of 100 gives $q=0.98$.
The distance $-\log q$ ($\infty$ when $q=0$) gives credit $q$, so general stability is the average fraction remaining.
For a user who tolerates a few replacements among the top 100, this distance describes the substantial agreement that the low top-100-set stability does not convey.
For six additional random targets fixed before sampling, the unchanged-set probability is only 0.084--3.63\%, yet 96.85--97.96 of the original top 100 remain on average.

  \paragraph{Priority distance reflects the importance of award decisions.}
Suppose a user cares more about preserving who receives an award than about preserving each recipient's award level.
Priority distance expresses this preference and ignores ranking changes that leave all awards unchanged.

 We illustrate this with the All-NBA structure: 15 players receive awards at three levels, five per level~\cite{nba2024allnba}.
Top-15 set distance ignores award levels; ordered top-15 distance counts swaps within a level even though the awards remain unchanged.
We use priority distance to give full credit when all awards are unchanged, half when only the recipients are unchanged, and zero otherwise.

 We rank the first 30 regular-season players using equal weights on five attributes and assign ranks 1--5, 6--10, and 11--15 to the three award levels.
We draw two million independent weights over the full positive weight space.
The same 15 players receive awards in 6.228\% of samples, while all award levels remain unchanged in 1.207\%.
General stability under the chosen $dist_{\rm pri}$ is 0.0372, combining agreement on recipients and award levels.

\subsection{  Accuracy of Stability Estimates }
\label{subsec:accuracy_check}

$GSC_{md}$, Prefix-GSC,  and Conv-SC are approximation methods: they do not return the ground-truth stability value, but an estimate whose accuracy is controlled by a user-specified error tolerance.
The theoretical analysis guarantees that, with high probability, the relative error is below the requested tolerance.
This guarantee  does not show what the approximation error looks like in practice, or how large the observed errors  are relative to the   tolerance.
We test both the estimation error and whether the returned confidence interval contains the reference value.

However, calculating the exact stability value is $\#P$-hard when dimension is part of the input (Section~\ref{sec:multi}).
We therefore use two kinds of experiments for which precise reference answers are available.
First, we take two or three attributes from real datasets and compute reference values by geometric integration.
Second, we construct multidimensional inputs whose stability values can be calculated directly from the construction.

\paragraph{Real data in low dimensions.}
We integrate the relevant regions of weight space to obtain reference values.
For exact stability, we test $GSC_{md}$ on 129 two-attribute queries with 30 items, using unscaled attributes, all attribute pairs, and random positive target weights.
These tests use QS 2026 data in place of QS 2023.
For $K_{0.5}$ and $F_{0.5}$, we use the first three attributes of 30-item QS, CWUR, and MPI inputs.
Prefix-GSC retains the first 11 and 12 positions, respectively.
Comparing its estimates with the full-distance reference values checks the combined error from sampling and leaving lower-ranked items unordered.
For Conv-SC, we test 24 priority-distance queries using two or three attributes from QS, CWUR, THE, and MPI.

\paragraph{Constructed multidimensional inputs.}
For $K_{0.5}$ and $F_{0.5}$, we construct inputs with two equally likely rankings that differ by exchanging the top two items.
Their general stability is the average contribution of these two rankings.
We test both $GSC_{md}$ and Prefix-GSC on 32 and 512 items through 256 attributes.

For Conv-SC, the coordinate-vector construction in Section~\ref{sec:not_poly} makes all $d!$ rankings equally likely.
Exact stability is therefore $1/d!$, and priority stability can be calculated by counting the rankings that preserve each group assignment.
We test Conv-SC's volume estimator for exact stability in four through six dimensions, and for priority distance in four, six, and eight dimensions.

Table~\ref{tab:accuracy-summary} reports the results for both kinds of experiments.
Every estimate meets the requested 2\% relative-error tolerance, and every returned confidence interval contains its reference value or integration interval.

\begin{table}[tbp]
\centering
\small
\textbf{All estimates meet the 2\% error target}\par\smallskip
\setlength{\tabcolsep}{2pt}
\begin{tabular}{@{}>{\raggedright\arraybackslash}p{0.45\linewidth}rrrr@{}}
\toprule
\textbf{Test} & \textbf{Trials} & \textbf{Dim.} & \textbf{Mean} & \textbf{Max.}\\
\midrule
\multicolumn{5}{@{}l}{\textit{Real data: geometric integration}}\\
$GSC_{md}$: exact & 6,450 & 2 & 0.30\% & 1.36\%\\
Prefix-GSC: $K_{0.5},F_{0.5}$ & 600 & 3 & 0.75\% & 1.91\%\\
Conv-SC: $dist_{\rm pri}$ & 72 & 2--3 & 0.43\% & 1.95\%\\
\midrule
\multicolumn{5}{@{}l}{\textit{Constructed inputs: known answers}}\\
$GSC_{md}$, Prefix-GSC: $K_{0.5},F_{0.5}$ & 288 & 16--256 & 0.23\% & 0.87\%\\
Conv-SC: exact & 300 & 4--6 & 0.26\% & 1.03\%\\
Conv-SC: $dist_{\rm pri}$ & 9 & 4, 6, 8 & 0.41\% & 0.93\%\\
\bottomrule
\end{tabular}
\caption{Accuracy on real data and constructed inputs: all $7{,}719$ estimates meet the 2\% error request.
Mean and maximum errors are absolute relative errors.}
\label{tab:accuracy-summary}
\end{table}

\subsection{When Computation Becomes Faster}
\label{subsec:cost}

Computing general stability can be exponentially expensive.
We test how the distance properties and methods in Section~\ref{sec:structured-computation} reduce this cost: efficient $GSC_{md}$ sampling under uniform boundedness, less sorting with Prefix-GSC, and convex-volume approximation with Conv-SC.

\begin{figure*}[tbp]
\centering
\includegraphics[width=\linewidth,alt={Two sampling panels occupy the left column. A vertical line separates them from eight method-comparison panels in two rows.}]{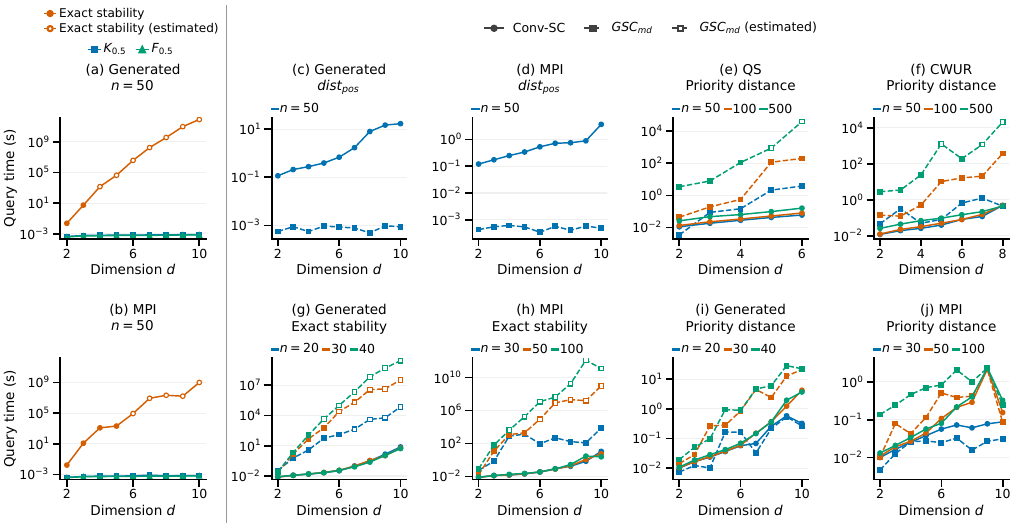}
\caption{Runtime as dimension grows, with 10\% relative error at 95\% confidence.
(a,b) $GSC_{md}$ remains fast for $K_{0.5}$ and $F_{0.5}$ while exact-stability queries become costly ($n=50$).
(c,d) $GSC_{md}$ is faster throughout the $dist_{pos}$ comparison.
(e--j) Conv-SC scales better for exact stability and several priority-distance settings.
Section~\ref{subsec:convsc_experiment} explains and checks the runtime estimates.}
\label{fig:sampling-scaling}
\label{fig:convsc-scaling}
\label{fig:convsc-exact-scaling}
\label{fig:distpos-scaling}
\Description{One figure with ten panels. Panels a and b in the left column compare distance choices for GSC-md on generated and MPI inputs. The eight panels to the right follow the discussion order: c and d show generated and MPI first-differing-position queries; e and f show QS and CWUR priority queries; g and h show generated and MPI exact queries; i and j show generated and MPI priority queries. The two groups have separate legends and share one caption. Hollow markers identify estimated runtimes.}
\end{figure*}

\subsubsection{\texorpdfstring{$GSC_{md}$}{GSCmd}: exponential worst case and polynomial cases}
\label{subsec:direct-experiment}

Uniform boundedness guarantees efficient sampling for $K_{0.5}$ and $F_{0.5}$ (Section~\ref{sec:fewer-draws}).
To see the practical difference, we compare $GSC_{md}$ under these distances and $0/\infty$ distance on the same randomly generated and MPI databases (Figure~\ref{fig:sampling-scaling}(a,b)).

Generated attributes combine a shared uniform $[0,1]$ score and independent standard Gaussian noise with weights $s_j=0.80-0.05j$ and $1-s_j$ for $j=1,\ldots,d$.
The shared score makes attributes partly agree on item quality, while the noise allows their rankings to differ.
Targets use random positive weights, and each point is a median over five generated inputs.
MPI uses the original attribute scales and equal target weights; position-weighted times are medians of five runs.

The runtime gap grows with dimension on both datasets.
On MPI, exact-stability queries take tens of minutes, while the position-weighted queries on the same inputs finish in under one millisecond.
Thus uniform boundedness gives a substantial computational benefit where exact-stability sampling is expensive.

  \paragraph{When both properties hold.}
$dist_{pos}$ is uniformly bounded and quasiconvex, so both $GSC_{md}$ and Conv-SC apply.
We compare them on 54 generated and MPI inputs with $n=50$ and $d=2$--10 (Figure~\ref{fig:distpos-scaling}(c,d)), using Section~\ref{sec:adaptive-interval}'s stopping rule.
We report three-run medians including setup, taking a further median over five inputs for each generated-data setting.
Since $dist_{pos}\leq1$, $GSC_{md}$ uses the known credit range $[e^{-1},1]$.
$GSC_{md}$ finishes in under 1.3 milliseconds and is faster on every input.
Thus quasiconvexity permits Conv-SC, while a small distance bound can make $GSC_{md}$ the faster choice.
In separate tests on 22 reference inputs, all 132 accuracy trials meet the error target; the maximum errors are 3.81\% for $GSC_{md}$ and 8.42\% for Conv-SC.

\subsubsection{Prefix-GSC reduces sorting work}
\label{subsec:prefix-experiment}

Prefix-GSC exploits tail decay with bounded influence to avoid fully sorting each sampled ranking.
We test whether this saves time overall, including setup and any extra samples needed to meet the same error tolerance.
We compare it with $GSC_{md}$ on the same generated databases, varying database size and dimension (Table~\ref{tab:weighted-scaling}).

\begin{table}[tbp]
\centering
\small
 \textbf{  Prefix-GSC speedup on generated data } \par\smallskip
\setlength{\tabcolsep}{3pt}
\begin{tabular*}{\linewidth}{@{\extracolsep{\fill}}rrrrrrr@{}}
\toprule
 &   \multicolumn{2}{c}{$d=6$}  &   \multicolumn{2}{c}{$d=32$}  &   \multicolumn{2}{c}{$d=128$} \\
  \textbf{Size} $n$  &   $K_{0.5}$  &   $F_{0.5}$  &   $K_{0.5}$  &   $F_{0.5}$  &   $K_{0.5}$  &   $F_{0.5}$ \\
  \midrule
500  &   1.40  &   1.51  &   1.23  &   1.22  &   1.08  &   1.08 \\
  2,000  &   1.81  &   2.07  &   1.37  &   1.39  &   1.11  &   1.11 \\
  10,000  &   \textbf{2.90}  &   \textbf{2.72}  &   1.52  &   1.55  &   1.16  &   1.16 \\
  \bottomrule
 \end{tabular*}
\caption{Prefix-GSC reduces total query time in every setting shown.
Values are median speedups over $GSC_{md}$, including setup and loading (three inputs, four paired trials each).}
  \label{tab:weighted-scaling}
 \end{table}

The sorting savings outweigh the extra sampling and setup in every setting in Table~\ref{tab:weighted-scaling}, with up to a $2.90\times$ speedup.
On 500-item QS, CWUR, and THE inputs with six, eight, and five attributes, respectively, the six queries show median speedups of $1.22$--$1.98\times$ over twelve paired trials per query.
All bootstrap lower bounds on speedup exceed 1.
Prefix-GSC retains 11 leading positions for $K_{0.5}$ and 12 for $F_{0.5}$, chosen before sampling by Section~\ref{sec:prefix-gsc}'s error bound.
For other position weights and distance scales, extra sampling can offset the sorting savings.

  \subsubsection{Conv-SC's scaling for quasiconvex distances}
 \label{subsec:convsc_experiment}

We test whether Conv-SC's runtime grows more slowly than $GSC_{md}$'s for exact stability and priority distance.
Its performance depends on how tightly the boxes enclose the regions (Section~\ref{sec:convex}); Theorem~\ref{thm:convsc-guarantee}'s polynomial guarantee assumes general volume routines.
Both methods request 10\% relative error at 95\% confidence.

\paragraph{Conv-SC's advantage for priority distance.}
QS uses ranking bands when score differences are too small to justify distinct ranks~\cite{qs_rank_bands}.
To test stability of such groups, we divide each target into six nearly equal groups and combine adjacent groups into three tiers.
Priority distance gives full credit if every item stays in its original group, half if all items stay in their original tiers but a group changes, and zero if any item changes tier.
We use 50, 100, and 500 universities and vary dimension from two to all available attributes: six for QS, five for THE, and eight for CWUR.
The 48 queries use equal target weights, normalization fixed on each 500-item dataset, and three trials per method.
$GSC_{md}$ uses the faster of comparison tests and full scoring, chosen by untimed pilot runs in its favor.

Conv-SC is $64\times$ faster for priority distance on 50 QS universities at six attributes (Figure~\ref{fig:convsc-scaling}(e)).
From two to six attributes, $GSC_{md}$'s measured runtime grows from 0.00343 to 3.79 seconds, about $1{,}100\times$, while Conv-SC grows from 0.0110 to 0.0594 seconds, only $5.4\times$.
Thirty further trials per method and dimension on QS-50 give growth factors of $790\times$ and $5.2\times$, respectively.
Across the nine dataset--size combinations, we compare runtime at all available attributes with runtime at two attributes.
Conv-SC has the smaller growth factor in eight cases, three using estimated $GSC_{md}$ endpoints (Figure~\ref{fig:convsc-scaling}(e,f)).
The exception is 50-item CWUR, where $GSC_{md}$ is faster at eight attributes.

  \paragraph{Exact stability and priority distance on other inputs.}
Figure~\ref{fig:convsc-scaling}(g--j) compares both distances on the same 162 generated and MPI inputs, using raw attribute scales.
Generated data use the attribute and target-weight distributions specified in Section~\ref{subsec:direct-experiment}.
We vary dimension from two to ten at database sizes 20, 30, and 40, reporting medians over five inputs per setting.
MPI uses Section~\ref{subsec:exp_1}'s row order, source-order attributes, and equal target weights (one input per setting).
For exact stability, $GSC_{md}$ checks adjacent comparisons until one fails.
Conv-SC completes all 162 exact queries; $GSC_{md}$ has 84 measured runtimes and 78 estimated runtimes.

  For $dist_{\rm pri}$, both methods complete all 162 queries, and Conv-SC is faster on 111.
On the 30- and 40-item generated databases, Conv-SC achieves speedups of about $30\times$ at seven attributes and $5$--$6\times$ at ten.
MPI does not show the same scaling advantage.
Its curves also reflect changes in the target ranking and group assignments as attributes are added.
Together with the university results, these experiments demonstrate Conv-SC's computational benefit for priority distance beyond exact stability.

  \paragraph{How accurate are the estimated runtimes?}
We predict the stopping checkpoint from independent estimates of the sample mean and variance, then use measured sampling and interval-checking costs to estimate runtime.
We complete runs with manageable predicted times and compare their actual runtimes with the estimates made beforehand.
Across 46 exact-stability queries and 22 priority-distance queries, the mean absolute relative errors of the runtime estimates are 2.9\% and 2.6\%, respectively.
We compare each estimate with the median measured runtime for that query.
The remaining hollow markers still represent estimates for unfinished runs.

\section{Related Work}

We extend ranking stability~\cite{asudeh2018obtaining} using a user-defined distance to compare rankings with a fixed target.
Multicriteria decision analysis studies choices and rankings based on several criteria~\cite{belton2002multiple}.
When the relative importance of these criteria is uncertain, stochastic multicriteria acceptability analysis explores possible weights and estimates how often an item attains a given rank~\cite{lahdelma1998smaa,lahdelma2001smaa2,tervonen2008survey}.
These probabilities describe the possible positions of individual items.
Our question concerns how much a reported ranking changes, with a user-defined distance specifying which changes matter.

Ranking methods also account for noisy or incomplete data by modeling item scores with probability distributions~\cite{li2010ranking}.
Work on uncertain rankings seeks rankings that minimize expected Kendall and Footrule distances~\cite{soliman2011uncertain}.
Local stability~\cite{campbell2026local} studies how changing a single item's attribute values affects its position, allowing swaps among items with similar quality.
We keep the data fixed and study changes in the scoring weights, comparing the resulting rankings with the reported ranking.
Standke et al.~\cite{standke2026parameters} use expected ranking changes
to study the importance of scoring parameters.

  Top-$k$ algorithms retrieve highest-scoring items
~\cite{fagin2001optimal,ilyas2008survey}; skyline and regret queries support
selection without fixing a scoring function
~\cite{borzsony2001skyline,nanongkai2010regret}.
Fairness constraints can specify how many items from a group appear in the top positions~\cite{celis2017ranking}.
Fair-ranking work finds nearby weight vectors satisfying fairness requirements~\cite{asudeh2019designing}.
These methods seek fair rankings, while we evaluate ranking robustness to changes in scoring weights.
Auditing work detects cherry-picked trendlines~\cite{asudeh2020detecting}
 and generalizations~\cite{lin2021detecting}.
These studies share our motivation of detecting conclusions that depend on a selectively chosen model.

Conv-SC builds on randomized algorithms that approximate convex-body volumes efficiently~\cite{kannan1997random,lovasz2006simulated}.
The connection to ranking stability comes from expressing ranking outcomes as regions in weight space.
Quasiconvexity lets Conv-SC combine all rankings within a distance threshold into one convex region.
This avoids approximating every ranking region separately when many rankings contribute to general stability.

\section{Conclusion}

We introduced general stability, a distance-based extension of exact ranking stability, to assess ranking changes according to their importance in an application.
We extended prior exact-stability methods to compute general stability and showed that uniform boundedness makes $GSC_{md}$ efficient despite its exponential worst case.
We also developed Prefix-GSC, an optimization that reduces sorting within $GSC_{md}$ under tail decay with bounded influence, and Conv-SC, which uses convex-volume approximation for quasiconvex distances.
Experiments on real and generated data show how distance choice affects ranking assessments and when these methods improve computation.
General stability lets users distinguish minor ranking changes from changes that affect the conclusions drawn from a ranking.

\clearpage
\balance
 \raggedbottom
 \bibliographystyle{ACM-Reference-Format}
\bibliography{sample}

\end{document}